\documentclass[twocolumn,showpacs,amsmath,amssymb,prl]{revtex4}

\usepackage{graphicx}   
\usepackage{dcolumn}    
\usepackage{bm}         
\usepackage{subfigure,textcomp}
\newcommand{\rd}{{\rm d}}
\newcommand{\re}{{\rm e}}
\newcommand{\ri}{{\rm i}}
\newcommand{\kB}{k_{\rm B}}
\newcommand{\vx}{\mathbf{r}}

\begin{document}

\title[Near-field heat transfer]{Near-field heat transfer in a scanning
       thermal microscope}

\author{Achim Kittel}
\author{Wolfgang M\"uller-Hirsch}
\author{J\"urgen Parisi}
\author{Svend-Age Biehs}
\author{Daniel Reddig}
\author{Martin Holthaus}


\affiliation{Institut f\"ur Physik, Carl von Ossietzky Universit\"at,
    D-26111 Oldenburg, Germany}

\date{August 31, 2005}

\begin{abstract}
We present measurements of the near-field heat transfer between the tip
of a thermal profiler and planar material surfaces under ultrahigh vacuum
conditions. For tip-sample distances below $10^{-8}$~m our results differ
markedly from the prediction of fluctuating electrodynamics. We argue that
these differences are due to the existence of a material-dependent small
length scale below which the macroscopic description of the dielectric
properties fails, and discuss a corresponding model which yields fair
agreement with the available data. These results are of importance for
the quantitative interpretation of signals obtained by scanning thermal
microscopes capable of detecting local temperature variations on surfaces.
\end{abstract}


\maketitle


Radiative heat transfer between macroscopic bodies increases strongly when
their spacing is made smaller than the dominant wavelength $\lambda_{\rm th}$
of thermal radiation. This effect, caused by evanescent electromagnetic fields
existing close to the surface of the bodies, has been studied theoretically
already in 1971 by Polder and van Hove for the model of two infinitely
extended, planar surfaces separated by a vacuum gap~\cite{PoldervanHove71},
and re-investigated later by Loomis and Maris~\cite{LoomisMaris94} and
Volokitin and Persson~\cite{VolokitinPersson01,VolokitinPersson04}. While early
pioneering measurements with flat chromium bodies had to remain restricted to
gap widths above~$1$~\textmu m~\cite{Hargreaves69}, and later studies employing
an indium needle in close proximity to a planar thermocouple remained
inconclusive~\cite{XuEtAl94}, an unambiguous demonstration of near-field heat
transfer under ultrahigh vacuum conditions and, thus, in the absence of
disturbing moisture films covering the surfaces, could be given in
Ref.~\cite{MuellerHirschEtAl99}.

The theoretical treatment of radiative near-field heat transfer is based
on fluctuating electrodynamics~\cite{RytovEtAl89}. Within this framework,
the macroscopic Maxwell equations are augmented by fluctuating currents
inside each body, constituting stochastic sources of the electric and
magnetic fields $\mathbf{E}$ and $\mathbf{H}$. The individual frequency
components $\mathbf{j}(\vx,\omega)$ of these currents are considered as
Gaussian stochastic variables. According to the fluctuation-dissipation
theorem, their correlation function reads~\cite{LL60}
\begin{equation}
\begin{split}
\langle j_\alpha(\vx,\omega) \, j_\beta^*(\vx',\omega') \rangle &\\
    & \hspace{-2cm} = \frac{\omega}{\pi}
    E(\omega,\beta) \, \epsilon'' \, (\omega) \, \delta_{\alpha\beta} \,
    \delta(\vx - \vx') \,  \delta(\omega - \omega') \; ,
\end{split}
\label{Eq:cor}
\end{equation}
where $E(\omega,\beta) = \hbar\omega/\big(\exp(\beta\hbar\omega) - 1\big)$,
with the usual inverse temperature variable $\beta = 1/(\kB T)$; the angular
brackets indicate an ensemble average. Moreover, $\epsilon''(\omega)$ denotes
the imaginary part of the complex dielectric function
$\epsilon(\omega) = \epsilon'(\omega) + \ri \epsilon''(\omega)$. It
describes the dissipative properties of the material under consideration,
which is assumed to be homogeneous and non-magnetic. Thus, Eq.~(\ref{Eq:cor}) contains
the idealization that stochastic sources residing at different points $\vx$,
$\vx'$ are uncorrelated, no matter how small their distance may be. Applied
to a material occupying the half-space $z < 0$, facing the vacuum in the
complementary half-space $z > 0$, these propositions can be evaluated to
yield the electromagnetic energy density in the distance $z$ above the
surface, giving~\cite{JoulainEtAl03}
\begin{equation}
\begin{split}
    \langle u(z) \rangle
      & = \frac{\epsilon_0}{2} \langle \mathbf{E}^2 \rangle
        + \frac{\mu_0}{2} \langle \mathbf{H}^2 \rangle  \\
      & = \int_0^\infty \!\!\!\!{\rd} \omega \int_0^\infty \!\!\!\! \rd \kappa \,
    \bigl(\rho_{\rm E}(\omega,\kappa,\beta,z) + \rho_{\rm H}(\omega,\kappa,\beta, z)
    \bigr) \\
      & = \int_0^\infty \!\!\!\! {\rd} \omega \,
        \frac{E(\omega,\beta) \omega^2}{2 \pi^2 c^3} \biggl\{ \!
        \int_0^1 \!\!\!\! {\rd} \kappa\, \frac{\kappa}{p}
    \big[1 + \kappa^2 {\rm Re}(r_\parallel {\re}^{2 {\ri} z  \omega p/c})\big] \\
      & \qquad\quad + \int_1^\infty \!\!\!\!{\rd} \kappa\, \frac{\kappa^3}{|p|}
        {\rm Im}(r_\parallel) {\re}^{-2 z \omega |p|/c}
        + X_\perp \biggr\} \; .
\end{split}
\label{Eq:dens}
\end{equation}
Here, the densities $\rho_{\rm E}$ and $\rho_{\rm H}$ symbolically specify the
electric and magnetic contribution, respectively; $r_\parallel$ denotes the
Fresnel amplitude reflection coefficient for TM-modes with wave vector of
magnitude $\omega\kappa/c$ parallel to the surface. The symbol $X_\perp$
abbreviates the corresponding terms for TE-modes. The wave vector oriented
normal to the surface, of magnitude $\omega p/c$, distinguishes propagating
modes with real $p = \sqrt{1 - \kappa^2}$ for $\kappa \le 1$ from evanescent
modes with imaginary $p = \ri \sqrt{\kappa^2 - 1}$ for $\kappa > 1$.

Expression~(\ref{Eq:dens}) for the energy density, obtained strictly within
the framework of macroscopic electrodynamics, diverges for small distances~$z$
from the surface; for $z/\lambda_{\rm th} \ll 1$, one finds the power law
$\langle u(z)\rangle \propto z^{-3}$~\cite{RytovEtAl89}. Hence, it has been
suggested that also the energy dissipated in the tip of a tiny probe close to
the surface should scale inversely proportional to the cube of the tip-sample
distance~\cite{Dorofeyev98,MuletEtAl01}. However, the entailing divergence
clearly is not borne out by the actual physics~\cite{Pan00,MuletEtAl01b,Pan01}.
The divergence may formally be avoided by replacing the upper boundary of
integration, $\kappa = \infty$ in Eq.~(\ref{Eq:dens}), by a finite cutoff
$\kappa_c$, thereby excluding the problematic large-wave number contributions
to the ``evanescent'' part of the energy density~\cite{VolokitinPersson04}. It
is important to note that the divergence of the energy density~(\ref{Eq:dens})
close to the material surface reflects a shortcoming of the underlying
macroscopic theory: Considering a metal, the dielectric properties of which
are largely determined by the conduction electrons, one expects that any
contributions from spatial Fourier components shorter than their mean free
path are inadequately dealt with~\cite{PoldervanHove71}. More generally, the
spatial delta-like correlation~(\ref{Eq:cor}) becomes problematic on length
scales such that the microscopic properties of the materials start to make
themselves felt. These observations, in their turn, imply that experiments on
fluctuating electromagnetic fields in the extreme near-field regime, where
traditional macroscopic fluctuating electrodynamics can no longer be taken
for granted, may yield important information on microscopic material
properties.

In this Letter, we report on measurements of the near-field heat transfer
between the tip of a scanning thermal microscope and surfaces of gold (Au)
or gallium nitride (GaN). We have fabricated a thermosensor, integrated
into the tip of a variable-temperature scanning tunneling microscope (VT-STM),
which allows us to determine the heat transfer even for tip-sample distances
on the order of $1$~nm. We argue that our sensor essentially probes the
near-field energy density close to the sample, and demonstrate that the
experimental data differ markedly from the standard prediction~(\ref{Eq:dens})
for distances below $10$~nm. A simple, but physically motivated ansatz for
the description of the short-range dielectric material properties then leads
to qualitative agreement with the measured data, allowing one to extract
material-dependent length scales~$L$ below which the macroscopic theory fails.


When assessing the near-field heat flux between two bodies of different
temperature, precise location of their positions of zero separation is
of key importance. Since no body has a mathematically flat surface, this is
to some extent a matter of definition. We have chosen to record the heat
transfer between a cooled sample and the tip of a VT-STM at nearly room
temperature, so that zero separation of the two surfaces corresponds to
a certain level of electron coupling, {\em i.e.\/}, to a certain tunnel
current. To exclude any mechanism of heat transfer other than radiation,
one has to work under ultrahigh vacuum (UHV) conditions. Otherwise,
any surface adsorbate, or surrounding gas, might result in additional
contributions to the heat transfer, masking the radiative effect.

\begin{figure}
  \centering
  \subfigure{\includegraphics[angle=0.0, height = 2.7cm]{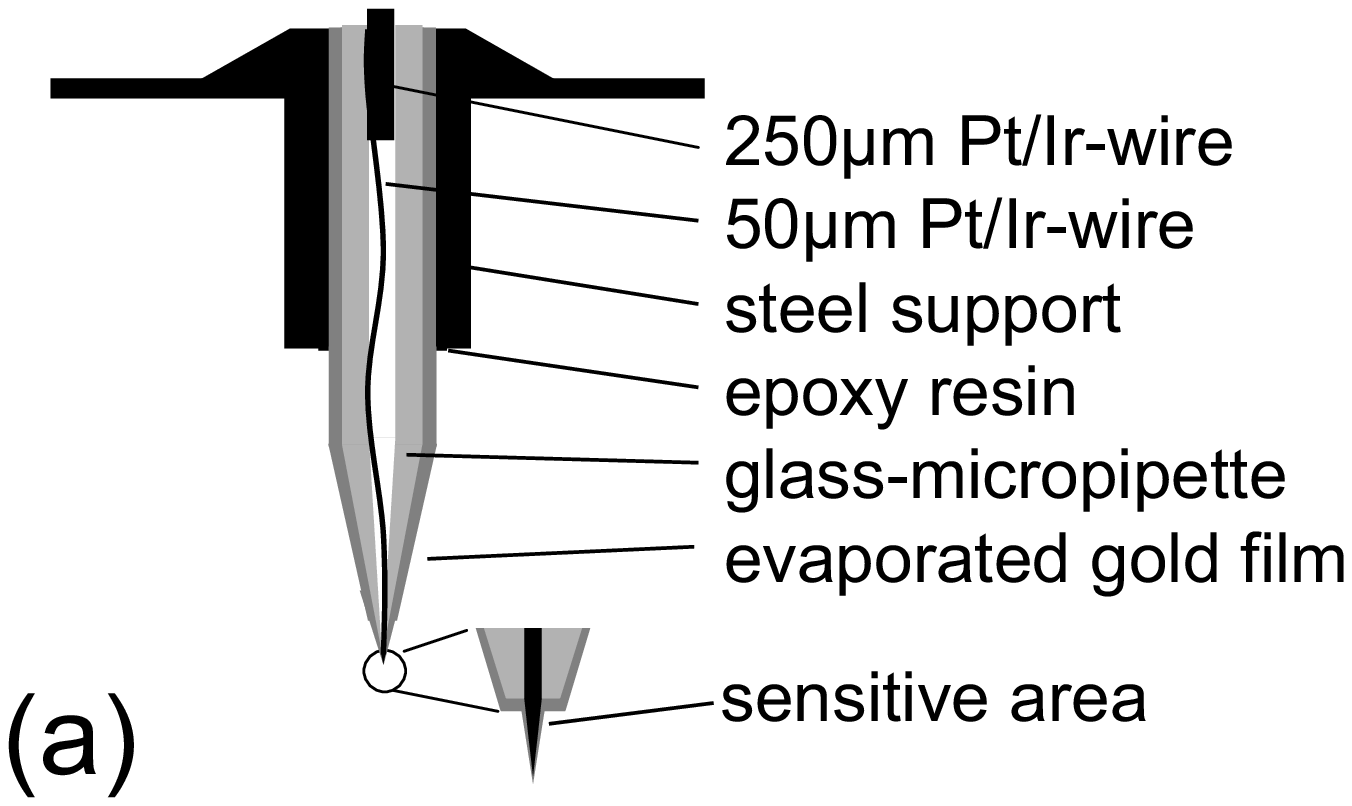}}
  \subfigure{\includegraphics[angle=0.0, height = 3.5cm]{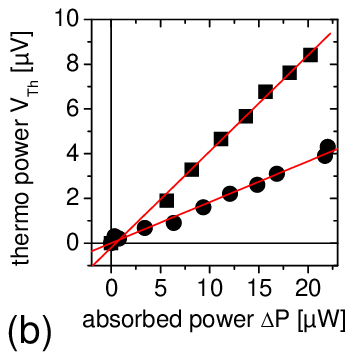}}
  \caption{(a) Cross section of the micropipette glued into a tip holder.
    The thermoelectric voltage $V_{\rm th}$ builds up between the inner
    platinum wire and the outer gold film. The tunnel potential is applied
    between the sample and the grounded gold film. (b) Dependence of the
    thermovoltage on the absorbed power $\Delta P$ of calibrating laser light
    for two different sensors.}
  \label{F_tip}
\end{figure}

The heat flux between the warm tip and the cooled sample is measured through
the resulting slight diminuation of the temperature of the very tip compared
to the rest of the sensor. Since small temperature differences have to be
detected over a small sensor, any self-heating has to be carefully avoided.
Therefore, we employ a thermocouple integrated into the tip of our VT-STM.
As sketched in Fig.~\ref{F_tip}~(a), a thin platinum wire has been melted
into a glass micropipette. Subsequently, the part of the wire protruding
from the pipette has been electrochemically etched to form a sharp
tip~\cite{LibioulleEtAl95}. The pipette has then been covered by a gold
film with a thickness of about $25$~nm, having electrical contact with the
platinum wire only at the very end of the tip. This end thus forms the
measuring contact of the resulting coaxial thermocouple, while the reference
contact is located in the back at the support of the micropipette, with good
thermal coupling to the surrounding which acts as a heat bath.

The support of the micropipette is sitting in the scanner of a commercial
VT-STM in an UHV chamber made by Omicron, held by a ring-shaped magnet.
In the hole of the magnet a gold-plated, spring-loaded contact pin makes
electrical contact with the platinum wire, thus forming the reference
contact. The gold film is grounded, whereas the sample is at tunnel potential,
in order to decouple the tunneling signal from the thermo\-electric
voltage~$V_{\rm th}$. This voltage is first amplified by a Keithley nanovolt
preamplifier (model 1801) and then measured by a high resolution multimeter
(Keithley model 2001). The temperature of the sample is lowered during the
measurements with liquid nitrogen via a coldfinger to $100$~K, establishing
a temperature difference between the tip and the sample surface of about
$200$~K.

The Seebeck coefficient $S = V_{\rm th}/\Delta T$~of our sensor, quantifying
the ratio of the generated thermoelectric voltage and the temperature
difference $\Delta T$ between the two contacts of the thermocouple, is
determined with a setup consisting of a droplet of oil held by a small heating
coil made of tungsten wire. The temperature of the oil is measured by a
commercial type-K thermocouple reaching from one side into the droplet, while
the sensor enters it from the other side. The oil temperature can be varied
by changing the current through the coil. We obtain $S = 8$~\textmu V/K at
room temperature, which is close to the value found in
literature~\cite{GotohEtAl91}.

\begin{figure}
\includegraphics[angle=0., scale=0.5, width = 7cm]{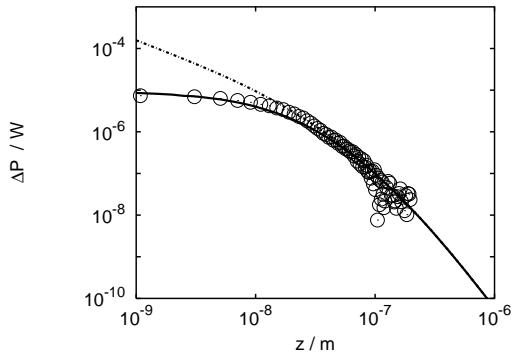}
\caption{Measured heat current $\Delta P$ (in Watts) between the microscope
    tip and a gold layer (circles) vs.\ tip-sample distance~$z$. The dashed
    line, which coincides with the full one for larger~$z$, corresponds to the
    prediction $\Delta P_{\rm th}$ of standard fluctuating electrodynamics,
    based on Eq.~(\ref{Eq:dens}). The full line is obtained from
    Eq.~(\ref{Eq:sab}) with the modified dielectric function~(\ref{Eq:mdf}),
    setting $L = L_{\rm tip} = 1.2 \cdot 10^{-8}$~m.}
\label{F_T1}
\end{figure}

\begin{figure}
\includegraphics[angle=0., scale=0.5, width = 7cm]{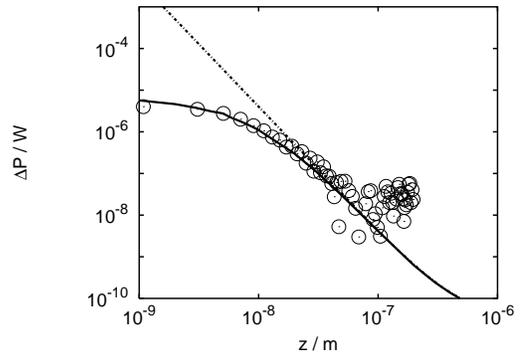}
\caption{As Fig.~\ref{F_T1} for a sample of GaN, setting
        $L =  1.0 \cdot 10^{-10}$~m and
    $L_{\rm tip} =  1.2 \cdot 10^{-8}$~m.}
\label{F_T2}
\end{figure}

The heat resistance $R_{\rm th} = \Delta T/\Delta P$ of the sensor, relating
the heat power~$\Delta P$ absorbed or emitted by the tip to the resulting
temperature difference between the contacts, was determined by placing the tip
in the focus of a $1$~mW cw laser diode (wavelength $670$~nm). The fraction of
the light power which did not hit the tip surface was measured by a power meter
positioned behind it. The absorbed power was estimated, according to
$\Delta P = (P_0-P) \cdot (1-R)$, from the difference $P_0 - P$ of the power
recorded without and with the tip being present, and the reflectivity $R=0.96$
of its gold surface at $670$~nm~\cite{BassEtAl85}. The expected linear
dependence of the thermovoltage on the absorbed power is well confirmed in
Fig.~\ref{F_tip}~(b), showing our results for two different tips. From the
slopes $0.18$~\textmu V/\textmu W and $0.43$~\textmu V/\textmu W we obtain
heat resistances of $23$~K/mW and $54$~K/mW, respectively. Knowing both a
sensor's Seebeck coefficient~$S$ and its heat resistance~$R_{\rm th}$,
one can deduce the near-field heat flux $\Delta P$ between the tip and a
closely spaced sample of different temperature from the observed thermovoltage,
according to $\Delta P = V_{\rm th} / \big(S R_{\rm th} \big)$. Measurements
of the distance dependence of the heat transfer were performed by retracting
the STM tip from the tunnel distance, while the distance itself was determined
by means of the piezo coefficient of the scanner. Results of such measurements
are depicted in Fig.~\ref{F_T1} for a sample consisting of a gold layer, and
in Fig.~\ref{F_T2} for a sample of GaN. In both cases, the sensor with
$R_{\rm th}= 54$~K/mW has been employed. During these measurements, we
have carefully checked that the crosstalk between the tunnel current signal
and the thermovoltage remains negligibly small. The absence of interference
is indicated by the fact that the tunnel current decreases strongly in a
range of distances where the observed thermovoltage stays almost constant.

A theoretical discussion of the heat transfer between an idealized tip and a
flat surface, which may serve as a guideline for the analysis of our data,
has been given by Mulet {\em et al.\/}~\cite{MuletEtAl01}. These authors have
modeled the tip by a small dielectric sphere of radius~$r$ and assumed the
incident electric field to be uniform inside the sphere, so that it acts
like a point-like dipole. If the temperature of the sample is significantly
lower than that of the tip, as in our case, the heat current flowing back
from the sample to the tip can be neglected. The total flux between the
surface and the tip then is determined entirely by the current directed
from the tip to the sample, according to
\begin{equation}
    \Delta P = \int_0^{\infty} \!\!\!\! \rd \omega
    \int_0^\infty \!\!\!\! \rd \kappa \,
    \alpha(\omega) \rho_{\rm E}(\omega,\kappa,\beta,z) \; ,
\label{Eq:dip}
\end{equation}
where
$\alpha(\omega) = 2 \omega (4 \pi r^3)
\epsilon_{\rm tip}''/|\epsilon_{\rm tip} + 2|^2$ describes the dielectric
properties of the sphere, and the temperature entering $\rho_{\rm E}$ is
that of the tip. Taking this expression at a representative
frequency~$\omega_0$, one has
$
    \Delta P \approx \alpha(\omega_0)
    \epsilon_0 \langle \mathbf{E}^2 \rangle/2 \; ,
$
so that, within the scope of the model, the heat flux registered by the tip
should be proportional to the electrical energy density of the flat sample,
evaluated, however, at the temperature of the tip.

For distances larger than about $10^{-8}$~m, our experimentally observed
heat transfer is, to good accuracy, proportional to the {\em total\/}
energy density as given by Eq.~(\ref{Eq:dens}), not to the electric field
contribution alone. Since the constant of proportionality, which carries
the dimension of area times velocity, may differ substantially from
$\alpha(\omega_0)$, we focus on the scaled energy density
$\Delta P_{\rm th} := \pi a^2 c \langle u(z) \rangle$, where $c$ is the
velocity of light, and employ the effective sensor area $\pi a^2$ as a
fitting parameter. Modeling the dielectric function $\epsilon(\omega)$ for
Au by a Drude ansatz with parameters taken from Ref.~\cite{AshcroftMermin76},
and that for GaN by the ``reststrahlen''-formula with parameters from
Ref.~\cite{Adachi04}, we obtain the dashed lines in Figs.~\ref{F_T1} and
\ref{F_T2}, setting $a = 60$~nm. This value is in accordance with
scanning electron microscopy studies of the tip, and describes {\em both\/}
experimental data sets for $z \gtrsim 10$~nm, as it should. The latter fact
also indicates that the use of Eq.~(\ref{Eq:dens}), {\em i.e.\/}, the neglect
of the field's distortion by the tip, is justified here.

In the case of GaN, the theoretical curve for $\Delta P_{\rm th}$ diverges
as $z^{-3}$ for sensor-sample distances below $10$~nm. In contrast, for
Au this familiar behaviour would become apparent only at substantially
smaller~$z$~\cite{VolokitinPersson01}. However, the experimental data clearly
show a different trend, leveling off to values which for the smallest
accessible distances are significantly lower than $\Delta P_{\rm th}$.
We interpret this finding as evidence for the short-distance deficiency
of the macroscopic theory, as expressed by the delta-like correlation
function~(\ref{Eq:cor}) of the stochastic source currents: In a real sample,
there is some finite correlation length~$L$.

In principle, one should then also account for non-local effects, which
requires distinguishing a transversal and a longitudinal part of the
permittivity~\cite{Ginzburg89}. Instead, here we propose a more simple,
qualitative approach: On the one hand, only the imaginary part
$\epsilon''(\omega)$ of the response function $\epsilon(\omega)$ enters
into the fluctuation-dissipation theorem and thus into the correlation
function~(\ref{Eq:cor}); on the other, the Kramers-Kronig formula relates
the imaginary part to the real one. Hence, a plausible and consistent ansatz
for an effective permittivity depending explicitly on the transversal wave
number is
\begin{equation}
    \widetilde{\epsilon}(\omega,\kappa) := 1 +
        \big[ \epsilon'(\omega) - 1 \big] f(\kappa) +
    \ri \epsilon''(\omega) f(\kappa) \; ,
\label{Eq:mdf}
\end{equation}
where the function $f(\kappa)$ accounts for the lateral correlations,
such that it approaches unity and thereby restores the local case when
$\omega\kappa/c \ll L^{-1}$, but vanishes for large wave numbers, when
$\omega\kappa/c \gg L^{-1}$. As a convenient guess, we take a Gaussian
$f(\kappa) = \exp\big(-(L \omega \kappa/c)^2\big)$, and consider $L$ as
a parameter to be determined by fitting the data. This
parametrization~(\ref{Eq:mdf}) has the distinct advantage that the Maxwell
equations for systems with plane translational invariance~\cite{PoldervanHove71}
remain formally unchanged; it is only that $\epsilon(\omega)$ has to be
replaced by $\widetilde{\epsilon}(\omega,\kappa)$. In particular, the energy
density can again be obtained from Eq.~(\ref{Eq:dens}), if only the reflection
coefficients $r_\parallel$ and $r_\perp$ are adapted in this manner.

Besides the dielectric properties of the sample, also those of the sensor
enter into the data, as exemplified by the dipole model~(\ref{Eq:dip}). Hence,
we have to introduce both a correlation length~$L$ of the sample and a further
correlation length $L_{\rm tip}$ of the sensor, and parametrize the
experimentally observed heat current in the form
\begin{equation}
    \frac{\widetilde{\Delta P}}{\pi a^2 c} =
    \int_0^\infty \!\!\! {\rd} \omega \! \int_0^\infty \!\!\!\! \rd \kappa \,
    \re^{- (L_{\rm tip} \omega\kappa/c)^2}
    \bigl(\rho_{\rm E}(\omega,\kappa) + \rho_{\rm H}(\omega,\kappa) \bigr)
    \; .
\label{Eq:sab}
\end{equation}
Using this ansatz (\ref{Eq:sab}), we finally obtain the full curves
in Figs.~\ref{F_T1} and \ref{F_T2}, setting
$L = 1.2 \cdot 10^{-8}$~m for Au
and $L = 1.0 \cdot 10^{-10}$~m for GaN, while
$L_{\rm tip} = 1.2 \cdot 10^{-8}$~m in both cases. These curves capture the
experimental data quite well, thus lending strong support to our line of
reasoning. It is also encouraging to observe that the numerical value of~$L$
obtained for Au indeed turns out to be on the order of the mean free path of
electrons in metals, whereas that for GaN is considerably shorter, as it
should. Although the thermally relevant component of our sensor probably is
confined to the Au layer, its correlation length not necessarily has to
coincide with that of the gold sample, as it actually does in our case, but
might be geometrically restricted in alternative setups.

In summary, we have obtained experimental data for the near-field heat
transfer between a thermal profiler and flat material surfaces under UHV
conditions. We have reached the extreme near-field regime, where the
variation of the heat transfer rate with the distance between microscope tip
and sample differs distinctly from the divergent behaviour predicted
by standard macroscopic fluctuating electrodynamics, and have interpreted
our observations in terms of finite microscopic correlations inside the
materials. While the shortcomings of the macroscopic theory are, in principle,
well known~\cite{RytovEtAl89,LL60}, their manifestation in an actual
experiment indicates a still unexplored potential of thermal microscopy as
a new, quantitative tool for the nanometer-scale investigation of solids.


\begin{thebibliography}{99}

\bibitem{PoldervanHove71} D. Polder and M. van Hove,
    Phys. Rev. B {\bf 4}, 3303 (1971).

\bibitem{LoomisMaris94} J. J. Loomis and H. J. Maris,
    Phys. Rev. B {\bf 50}, 18517 (1994).

\bibitem{VolokitinPersson01} A. I. Volokitin and B. N. J. Persson,
    Phys. Rev. B {\bf 63}, 205404 (2001).

\bibitem{VolokitinPersson04} A. I. Volokitin and B. N. J. Persson,
    Phys. Rev. B {\bf 69}, 045417 (2004).

\bibitem{Hargreaves69} C. M. Hargreaves,
    Phys. Lett. {\bf 30 A}, 491 (1969).

\bibitem{XuEtAl94} J.-B. Xu, K. L\"auger, R. M\"oller, K. Dransfeld,
    and I. H. Wilson,
    J. Appl. Phys. {\bf 76}, 7209 (1994).

\bibitem{MuellerHirschEtAl99} W. M\"uller-Hirsch, A. Kraft, M. Hirsch,
    J. Parisi, and A. Kittel,
    J. Vac. Sci. Technol. A {\bf 17}, 1205 (1999).

\bibitem{RytovEtAl89} S. M. Rytov, Yu. A. Kravtsov, and V. I. Tatarskii,
    {\em Principles of Statistical Radiophysics\/}
    (Springer, New York, 1989), Vol.~3.

\bibitem{LL60} E. M. Lifshitz and L. P. Pitaevskii,
    {\em Statistical Physics, Part~2\/}
    (Landau and Lifshitz: Course of Theoretical Physics, Vol.~9;
    Butterworth-Heinemann, Oxford, 2002).

\bibitem{JoulainEtAl03} K. Joulain, R. Carminati, J.-P.\ Mulet, and
    J.-J.\ Greffet,
    Phys. Rev. B {\bf 68}, 245405 (2003).

\bibitem{Dorofeyev98} I. A. Dorofeyev,
    J. Phys. D: Appl. Phys. {\bf 31}, 600 (1998).

\bibitem{MuletEtAl01} J.-P.\ Mulet, K. Joulain, R. Carminati, and
    J.-J.\ Greffet,
    Appl. Phys. Lett. {\bf 78}, 2931 (2001).

\bibitem{Pan00} J. L. Pan,
    Opt. Lett. {\bf 25}, 369 (2000).

\bibitem{MuletEtAl01b} J.-P.\ Mulet, K. Joulain, R. Carminati, and
    J.-J.\ Greffet,
    Opt. Lett. {\bf 26}, 480 (2001).

\bibitem{Pan01} J. L. Pan,
    Opt. Lett. {\bf 26}, 482 (2001).

\bibitem{LibioulleEtAl95} L. Libioulle, Y. Houbion, and J. M. Gilles,
    Rev. Sci. Instr. {\bf 66}, 97 (1995).

\bibitem{GotohEtAl91} M. Gotoh, K. D. Hill, and E. G. Murdock,
    Rev. Sci. Instr. {\bf 62}, 2778 (1991).

\bibitem{BassEtAl85} J. Bass, J. Dugdale, C. Foiles, and A. Myers,
    {\em Landolt-B{\"o}rnstein; Metals: Electronic Transport Phenomena}
    {\bf III/15b}, {\em Electrical Resistivity, Thermoelectrical Power
    and Optical Properties\/} (Springer, Berlin, 1985).

\bibitem{AshcroftMermin76} N. W. Ashcroft and N. D. Mermin,
    {\em Solid State Physics\/}
    (Harcourt, Fort Worth, 1976).

\bibitem{Adachi04} S. Adachi,
    {\em Handbook on Physical Properties of Semiconductors\/}
    (Kluwer, Boston, 2004).

\bibitem{Ginzburg89} V. Ginzburg,
    {\em Applications of Electrodynamics in Theoretical Physics
    and Astrophysics\/}
    (Gordon and Breach, New York, 1989).

\end{thebibliography}
\end{document}